\documentclass[aps,prl,twocolumn,times,graphicx,amssymb,tighten,showpacs,groupedaddress,superscriptaddress]{revtex4-1}

\usepackage{bm}
\usepackage{amsmath}
\usepackage{graphicx}
\usepackage{graphics}
\usepackage[dvips]{color}

\begin{document}

\title{Relativistic descriptions of final-state interactions
in charged-current quasielastic neutrino-nucleus scattering at MiniBooNE kinematics.}
\author{Andrea Meucci} 
\affiliation{Dipartimento di Fisica Nucleare e Teorica, 
Universit\`{a} degli Studi di Pavia and \\
INFN, Sezione di Pavia, Via A. Bassi 6, I-27100 Pavia, Italy}
\author{M.B. Barbaro} 
\affiliation{Dipartimento di Fisica Teorica, Universit\`a di Torino and
  INFN, Sezione di Torino, Via P. Giuria 1, 10125 Torino, Italy}
\author{J.A. Caballero}
\affiliation{Departamento de F\'{\i}sica At\'{o}mica, Molecular y Nuclear,
Universidad de Sevilla,
  41080 Sevilla, Spain}
\author{C. Giusti}
\affiliation{Dipartimento di Fisica Nucleare e Teorica, 
Universit\`{a} degli Studi di Pavia and \\
INFN, Sezione di Pavia, Via A. Bassi 6, I-27100 Pavia, Italy}
\author{J.M. Ud\'{\i}as}
\affiliation{Grupo de F\'{\i}sica Nuclear, Departamento de
F\'{\i}sica At\'{o}mica, Molecular y Nuclear,
Universidad Complutense de Madrid, CEI Moncloa,
  28040 Madrid, Spain}

%\date{\today}

\begin{abstract}

The results of two relativistic models with different descriptions of the 
final-state interactions are compared with the MiniBooNE data of 
charged-current quasielastic cross sections. The relativistic mean field model 
uses the same potential for the bound and ejected nucleon wave functions. In 
the relativistic Green's function (RGF) model the final-state interactions are 
described in the inclusive scattering consistently with the exclusive scattering 
using the same complex optical potential. 
The RGF results describe the experimental data for total cross-sections without 
the need to modify the nucleon axial mass.
\end{abstract}

\pacs{ 25.30.Pt; 13.15.+g; 24.10.Jv}

\maketitle

The double differential cross sections for muon neutrino charged-current 
quasielastic (CCQE) scattering, recently measured by the MiniBooNE 
collaboration~\cite{miniboone}, have raised debate over the role of the various 
theoretical ingredients entering the description of the reaction. 
High-quality descriptions of the CCQE differential cross sections in the 
few-GeV region are required to support neutrino oscillation 
measurements~\cite{miniboone,NOMAD}. The energy region explored requires a 
relativistic description of the process, where not only relativistic kinematics 
is considered, but also nuclear dynamics and current 
operators should be described within a relativistic 
framework. 

The simplest relativistic model to describe CCQE neutrino scattering is the 
relativistic Fermi Gas (RFG). When a dipole shape is assumed for the axial form 
factor, the nucleon axial mass $M_A$ has been used as a free parameter within 
the RFG model. 
Indeed, the MiniBooNE cross section~\cite{miniboone} is underestimated by the RFG unless $M_A$ is significantly enlarged (1.35 GeV/$c^2$) with respect to the 
accepted world average value (1.03 GeV/$c^2$~\cite{Bern02}).
As it turns out from comparison with electron scattering data, the RFG is too crude to correctly account for the nuclear  dynamics. Thus, a larger axial mass within the RFG could be a way to effectively incorporate nuclear effects.

More sophisticated models have been applied to
neutrino-nucleus scattering. At the level of the impulse approximation (IA), 
models based on a realistic spectral function~\cite{ben10} or on the 
relativistic IA (RIA) which, contrarily to RFG, are in good agreement with 
electron scattering data, also underestimate the experimental CCQE cross 
sections~\cite{Butkevich10,amaro11a,amaro11b} unless $M_A$ is significantly
enlarged, as indicated by the RFG prediction. 

It has been pointed out that in some kinematic regions where the neutrino flux 
for the experiment has significant strength, the reaction may have sizable 
contributions from effects beyond the IA. 
These include two-particle-two-hole (2p-2h) excitations, which may be reached 
via two-body meson-exchange currents (MEC). 
The contribution of the vector MEC in the 2p-2h sector, evaluated in the model 
of Ref.~\cite{De Pace:2003xu}, has been incorporated in a phenomenological 
approach, indicated as SuSA, based on the SuperScaling behavior of electron 
scattering data~\cite{amaro11a}. The strict SuSA predictions show a systematic 
discrepancy in comparison with the MiniBooNE cross sections~\cite{amaro11a}. 
The inclusion of 2p-2h MEC contributions yields 
somewhat better agreement with the data~\cite{amaro11b}, although theory still 
lies below the data at larger angles. 
Other theoretical results that incorporate multiple knockout excitations are 
in accordance with the experimental cross sections without the need to increase 
the value of $M_A$~\cite{Nieves11,Martini}. We note that the results 
in~\cite{Nieves11} are obtained in a relativistic  approach and in 
\cite{Martini} in a non-relativistic one.  

Within the QE kinematic domain, the treatment of the final-state interactions 
(FSI) between the ejected nucleon and the residual nucleus has been proved to 
be essential to compare to data. The relevance of FSI has been clearly 
stated for {\em exclusive} $(e,e^{\prime}N)$ processes, where the use of 
complex optical potentials in the distorted-wave impulse approximation (DWIA) 
is required~\cite{book,Ud1,meucci1}. In the analysis of {\em inclusive} 
reactions, FSI remains a crucial ingredient for a proper description 
of data~\cite{Chiara03,cab1,amaro05,cc,ee,confee,cab10}. All elastic and 
inelastic channels contribute to the inclusive process. Thus, the complex 
potential, with imaginary terms designed to reproduce just the elastic channel, 
should be dismissed. 
Different approaches have been used to account for FSI under inclusive 
conditions. For instance, in the approaches based on the relativistic DWIA (RDWIA), 
they have been accounted for by using purely real potentials. The final nucleon 
state has been evaluated with the real part of the relativistic 
energy-dependent optical potential (rROP), or with the same relativistic mean 
field potential considered in describing the initial nucleon state 
(RMF)~\cite{Chiara03,cab1}. However, the rROP is unsatisfactory from a 
theoretical point of view, since it is an energy-dependent potential, 
reflecting the different contribution of open inelastic channels for each 
energy, and under such conditions dispersion relations dictate that the 
potential should have a nonzero imaginary term~\cite{hori}. 
On the other hand, the RMF model is based on the use of the same strong 
energy-independent real potential for both bound and scattering states. 
It fulfills the dispersion relations~\cite{hori} and also the continuity 
equation. The RMF model applied to inclusive QE $(e,e')$ processes describes 
scaling behavior and gives rise to a superscaling function with a significant 
asymmetry, in good agreement with data~\cite{cab1,cab07}.

A different description of FSI makes use of relativistic
Green's function techniques~\cite{cc,ee,eenr,confee}.
Under suitable approximations~\cite{cc,ee,eenr,eeann,Meucci:2005pk}, that are 
basically related to the IA, the components of the nuclear response are written 
in terms of the single-particle (s.p.) optical model Green's function. This 
formalism allows to recover the contribution of non-elastic channels in the case 
of inclusive scattering, starting from the complex relativistic optical 
potential (ROP) which describes elastic nucleon-nucleus scattering data.
The relativistic Green's function (RGF) model allows for a consistent treatment 
of FSI in the exclusive and in the inclusive scattering and gives also a good 
description of $(e,e')$ data~\cite{ee,confee}. 

The results of the RMF and RGF models have been compared for the inclusive QE 
electron scattering~\cite{confee} and for the CCQE neutrino 
scattering~\cite{Meucci11}. As mentioned, both models describe successfully the 
behavior of electron scattering data and their scaling and superscaling 
functions and both produce a significant asymmetry in the scaling function that 
is strongly supported by data. There are, however, some differences between the 
RMF and RGF results depending on kinematics, which increase with the momentum 
transfer. Whereas the RMF 
%These differences can be understood if one considers than the RMF 
may be considered as a faithful representation of the pure ``nucleonic'' contribution to the 
inclusive response, the RGF, on the contrary, may to some extent translate loss of elastic 
strength to non-nucleonic degrees of freedom, contributing to the imaginary 
optical potential, into inclusive strength predicted by the RGF.

In this letter the predictions of the RMF and RGF models are compared with the 
recent CCQE MiniBooNE data.
The comparison between the results of the two models~\cite{confee,Meucci11}, 
which make use of very different ingredients, can be helpful for a deeper 
understanding of nuclear effects, more specifically FSI, which may play a 
crucial role in the analysis of CCQE data and its influence in studies of 
neutrino oscillations at intermediate to high energies.
This is of particular interest for the case of the MiniBoone CCQE data which, 
given the nature of the experiment, may receive more than pure nucleonic 
contributions~\cite{Tina10}. Thus the RMF would represent a lower bound to 
MiniBoone CCQE, while the RGF
%, which may receive non-nucleonic contributions, 
should yield larger predictions.

Details of the two models can be found 
in~\cite{Chiara03,cab1,cab07,cab05,amaro07b} for the RMF and 
in~\cite{cc,ee,eenr,eeann,Meucci:2005pk,Meucci11,Meucci06,Meucci:2006cx} for 
the RGF. In the RMF case, the components of the nuclear response are obtained from the sum over all the s.p. shell-model states of the squared absolute value of the transition matrix elements of the single-nucleon current. In the RGF case, the calculations require matrix elements involving the 
eigenfunctions of a complex optical potential and of its Hermitian conjugate~\cite{cc,ee}. 

In both calculations the bound nucleon states 
are self-consistent Dirac-Hartree solutions derived 
within a RMF approach using a Lagrangian containing $\sigma$, $\omega$, and 
$\rho$ mesons~\cite{bound}. 
The same real potential gives the scattering states in the RMF, whereas in the 
RGF calculations two parameterizations for the ROP have been used: the 
energy-dependent and A-dependent EDAD1 and the energy-dependent but 
A-independent EDAI-12C complex phenomenological potentials of~\cite{chc}, which 
are fitted to proton elastic scattering data on several nuclei in an energy 
range up to 1040 MeV. 
The comparison between the results obtained with two different phenomenological 
optical potentials may indicate how the incomplete determination of this 
important ingredient can influence the predictions of the model. 
In all the calculations we have used the standard value of the nucleon axial 
mass, {\it i.e.}, $M_A = 1.03$ GeV/$c^2$.
 
%%%%%%%%%%%%%%%%%%%%%%%%%%%%%%%%%%%%%%%%%%%%%%%%%%%%%%%%%%
\begin{figure}[ht]
%\label{fig:1}
\includegraphics[scale=0.4]{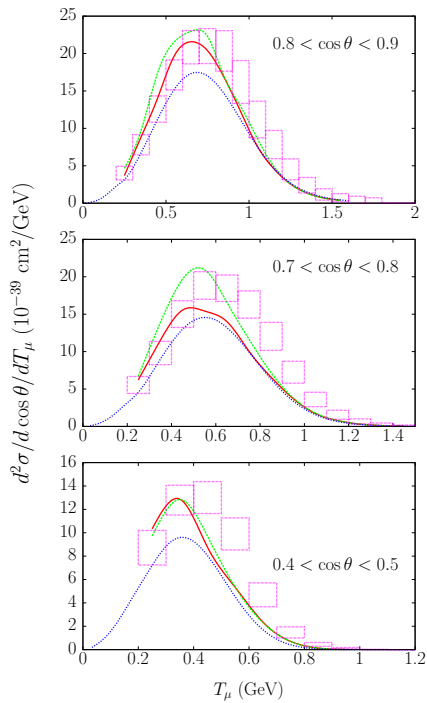}%
%\includegraphics[scale=0.4]{fig1.epsi}%
%\includegraphics[scale=0.4]{COS085.epsi}%
%\\
%\includegraphics[scale=0.4]{COS075.epsi}%
%\\
%\includegraphics[scale=0.4]{COS045.epsi}%
%
\vskip -0.3cm
\caption{(color online) Flux-averaged double differential cross section per target nucleon for the CCQE $^{12}$C$(\nu_{\mu} , \mu ^-)$ reaction calculated in the RMF (blue line) and in the RGF  with EDAD1 (red) and 	EDAI (green) potentials and displayed versus $T_\mu$ for various bins 	of $\cos\theta$. The data are from MiniBooNE~\cite{miniboone}. 
The uncertainties do not include the overall normalization error $\delta N$=10.7\%. 
\label{f1}}
\end{figure}
%%%%%%%%%%%%%%%%%%%%%%%%%%%%%%%%%%%%%%%%%%%%%%%%%%%%%%%%
In Fig.~\ref{f1} we show the CCQE double-differential 
$^{12}$C $(\nu_{\mu},\mu^{-})$ cross section averaged over the neutrino flux as a function of the muon kinetic energy $T_\mu$. In each panel the results have been averaged over the corresponding angular bin of $\cos\theta$, where $\theta$ is the scattering angle of the muon. The results evaluated with RMF (blue line) and RGF with EDAD1 (red) and EDAI (green) potentials are compared with the MiniBooNE CCQE data~\cite{miniboone}.   

The RMF results~\cite{amaro11b} yield reasonable agreement with data for small angles and low muon energies, the discrepancy becoming larger as $\theta$ and $T_\mu$ increase. The shape followed by the RMF cross sections fits well the slope shown by the data. A good agreement with the experimental shape is shown also by the RGF cross sections. The RMF and RGF models yield close predictions at larger values of $T_\mu$ for all the bins of $\cos\theta$ shown in the figure. 
Notice, however, that the RGF cross sections are generally larger than the RMF ones, particularly around the peak region, where the RGF produces cross sections in reasonable agreement with data.

It is worth noticing that the differences between the RGF results obtained with 
the two optical potentials are enhanced in the peak region and are in general 
of the order of the experimental errors. The EDAD1 and EDAI potentials yield
close predictions for the bin $0.4<\cos\theta<0.5$, a small differences is seen 
in the bin $0.8<\cos\theta<0.9$, being the RGF-EDAI cross section larger than 
the RGF-EDAD1 one, while the difference is sizeable for the bin 
$0.7<\cos\theta<0.8$, with the RGF-EDAD1 results closer to the RMF than to the 
RGF-EDAI ones.

The RMF model uses the effective mean field, that reproduces the saturation behavior of nuclear 
matter and the properties of the ground state of nuclei. It includes only nucleonic contributions 
to the inclusive process. The RGF uses 
phenomenological optical potentials, fitted to elastic proton-nucleus scattering. The loss of 
elastic flux into inelastic channels (either multi-nucleon knockout as well as non nucleonic 
excitations) caused by the imaginary term of these potentials is recovered for the inclusive 
scattering making use of dispersion relations. The larger cross section shown by the RGF 
%yields a larger cross section than the RMF has to 
can be attributed to non purely nucleonic inelasticities represented 
in the phenomenological ROP~\cite{confee,Meucci11}.

%%%%%%%%%%%%%%%%%%%%%%%%%%%%%%%%%%%%%
\begin{figure}[ht]
\includegraphics[scale=0.4]{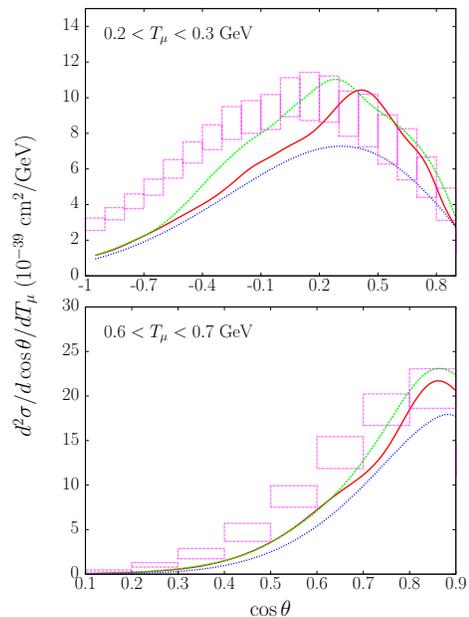}%
%\includegraphics[scale=0.4]{fig2.epsi}%
%\label{fig:2}
%\includegraphics[scale=0.4]{T025_smooth.epsi}%
%\\
%\includegraphics[scale=0.4]{T065_smooth.epsi}
\vskip -0.3cm
\caption{(color online)  Flux-averaged double differential cross section per 
target nucleon for the CCQE $^{12}$C$(\nu_{\mu} , \mu ^-)$ reaction displayed
versus $\cos\theta$ for two bins of $T_\mu$. The results obtained with 
RMF (blue line), RGF EDAD1 (red), and RGF EDAI (green) potentials are compared
with the MiniBooNE data of~\cite{miniboone}. \label{f2}}
\end{figure}
%%%%%%%%%%%%%%%%%%%%%%%%%%%%%%%%%%%%%

In Fig.~\ref{f2} the flux-averaged double differential cross sections are 
plotted versus $\cos\theta$ for two bins of $T_\mu$, {\it i.e.}, 
$0.2<T_\mu<0.3$ GeV and  $0.6<T_\mu<0.7$ GeV. The approximate shape of the 
experimental cross section is well described by the models. The RMF results 
generally underestimate the data, especially for the lower muon energy values, 
the agreement improves as $T_\mu$ increases. The RGF provides a better accordance
with the size of the experimental cross section. The agreement is 
better for smaller angles while the data are slightly underpredicted as 
$\theta$ increases. The RGF-EDAD1 yields in general a lower cross section than 
the RGF-EDAI, yet higher than the RMF one.

%%%%%%%%%%%%%%%%%%%%%%%%%%%%%%%%%%
\begin{figure}[ht]
%\label{fig:3}
\includegraphics[scale=0.5]{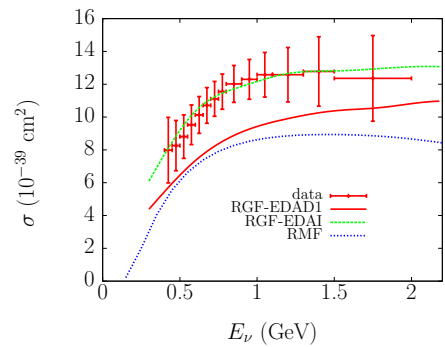}%
\vskip -0.3cm
\caption{(color online) Total CCQE cross section per neutron versus the neutrino energy. The cross sections calculated in the RMF (blue line), RGF EDAD1 (red), and RGF EDAI (green) potentials are compared with the flux unfolded MiniBooNE 
data of~\cite{miniboone}.
 \label{f3}.}
\end{figure}
%%%%%%%%%%%%%%%%%%%%%%%%%%%%%%%%%%

Finally, in Fig.~\ref{f3} the total QE cross section per neutron obtained in 
the RMF and RGF models are displayed as a function of the neutrino energy 
$E_\nu$ and compared with
the ``unfolded'' experimental data~\cite{miniboone}. 
It was shown in~\cite{amaro11b} that the differences between the results of 
the RMF, SuSA, and rROP models tend to be washed out in the integration and 
that all these models, representing essentially the same nucleonic contribution 
to the inclusive cross sections, yield very similar results, all of them 
underpredicting the total MiniBoone CCQE experimental cross section. Larger 
cross sections, in particular for larger values of $E_\nu$, are obtained in the 
RGF with both optical potentials. The differences between RGF-EDAI and RGF-EDAD1 
are here clearly visible, being RGF-EDAI in good agreement with the shape and 
magnitude of the experimental cross section and RGF-EDAD1 above RMF but clearly 
below the data. The differences between EDAI and EDAD1 are due to the
different values of the imaginary parts of both potentials, particularly for 
the energies considered in kinematics with the lowest $\theta$ and the largest 
$T_\mu$. These kinematics, which were not considered in previous RGF 
calculations, give large contributions to the total cross section and emphasize 
the differences between the RGF predictions with both optical potentials. 
Notice that EDAI is a single-nucleus parameterization, which does have an edge 
in terms of better reproduction of the elastic proton-$^{12}C$ phenomenology 
\cite{chc} compared to EDAD1, and also leads to CCQE results in better 
agreement with data.

Summarizing, in this letter the results of the RMF and RGF models have been 
compared with the recent CCQE MiniBooNE data. 
Both models give a good description of the shape of the experimental cross 
sections. The RMF generally underpredicts the data, particularly for lower 
values of $\theta$ and $T_\mu$. In contrast, the RGF  can give cross 
sections of the same magnitude as the experimental ones without the need to 
increase the standard value of the axial mass. The larger cross sections in the 
RGF model arise from the translation to the inclusive strength of the overall 
effect of inelastic channels (nucleonic and non-nucleonic). At present, lacking 
a phenomenological optical potential which exactly fullfills the dispersion 
relations in the whole energy region of interest, the RGF prediction is not 
univocally determined from the elastic phenomenology, though some preference to 
the EDAI predictions should be given.
 
Our results give a further and clear indication that before drawing conclusions 
about the comparison with data, a careful evaluation of all nuclear effects and 
the possible effect of some non-nuclenic contributions to CCQE MiniBoone data 
is required \cite{Tina10}. This is important also to reconcile former results 
for which RMF was in good agreement with previous CCQE data~\cite{Cris06} 
A better determination of a phenomenological relativistic optical potential 
which closely fullfills the dispersion relations deserves further investigation.

This work was partially supported by DGI (Spain) under contract nos. FIS2008-04189, FPA2010-17142, the Spanish Consolider-Ingenio 2010 programme CPAN (CSD2007-00042), and by the INFN-CICYT collaboration agreements ACI2009-1053 and AIC10-D-000571.

%%%%%%%%%%%%%%%%%%%%%%%%%%%%%%%%%%%%%%%%%%%%%%%%%%%%%%%%%%%%%%%%%%%%
%


\begin{thebibliography}{99}
%
\bibitem{miniboone}
A.~A. Aguilar-Arevalo, {\it et al.} [MiniBooNE Collaboration],
Phys. Rev. D {\bf 81}, 092005 (2010).
%
\bibitem{NOMAD}
V. Lyubushkin, {\it et al.}, Eur. Phys. J. C
{\bf 63}, 355 (2009); 
P. Adamson, {\it et al.}, 
Phys. Rev. D {\bf 81}, 072002 (2010).
%
\bibitem{Bern02}
 V. Bernard, L. Elouadrhiri, and U.G. Meissner, J. Phys. G {\bf 28}, R1 (2002);
% V. Bernard, {\it et al.}, J. Phys. G {\bf 28}, R1 (2002);
A. Bodek, {\it et al.}, Eur. Phys. J. C {\bf 53}, 349
(2008).
\bibitem{ben10}
O. Benhar, P. Coletti, and D. Meloni, Phys. Rev. Lett. {\bf 105}, 132301 (2010); 
C. Juszczak, J.~T. Sobczyk, and J. Zmuda, Phys. Rev.  {\bf 82}, 045502 (2010);
%
\bibitem{Butkevich10}
  A.~V.~Butkevich,
  Phys.\ Rev.\  C {\bf 82}, 055501 (2010).
%
\bibitem{amaro11a}
J.~E. Amaro, {\it et al.}, 
Phys. Lett.  {\bf B696}, 151 (2011).
%
\bibitem{amaro11b}
J.~E. Amaro, {\it et al.}, 
[arXiv:1104.5446 [nucl-th]].
%
\bibitem{De Pace:2003xu}
  A.~De Pace,  {\it et al.},
  %M.~Nardi, W.~M.~Alberico, T.~W.~Donnelly and A.~Molinari,
  %``The 2p - 2h electromagnetic response in the quasielastic peak and beyond,''
  Nucl.\ Phys.\  A {\bf 726} (2003) 303.
  %%CITATION = NUPHA,A726,303;%%
%
\bibitem{Amaro:2010iu}
  J.~E.~Amaro, C.~Maieron, M.~B.~Barbaro, J.~A.~Caballero and T.~W.~Donnelly,
  %``Pionic correlations and meson-exchange currents in two-particle emission
  %induced by electron scattering,''
  Phys.\ Rev.\  C {\bf 82} (2010) 044601.
%  [arXiv:1008.0753 [nucl-th]].
  %%CITATION = PHRVA,C82,044601;%%
%
\bibitem{Nieves11}
  J.~Nieves, I. Ruiz Simo,  and M.~J.~Vicente Vacas,
  Phys.\ Rev.\  C {\bf 83}, 045501 (2011); [arXiv:1106.5374 [nucl-th]].
%
\bibitem{Martini}
  M. Martini, M. Ericson, G. Chanfray, and J.Marteau,
  Phys.\ Rev.\  C {\bf 80}, 065501 (2009);
  Phys.\ Rev.\  C {\bf 81}, 045502 (2010).
%
%\bibitem{depace03}
%A. De Pace {\it et al.},
%Nucl. Phys.  {\bf A726}, 303 (2003). 
%
\bibitem{book}
S. Boffi, C. Giusti, F.D. Pacati, and M. Radici,
{\it Electromagnetic Response of Atomic Nuclei}, Oxford Studies in Nuclear
Physics, Vol. 20 (Clarendon Press, Oxford, 1996);
S. Boffi, C. Giusti, and F.D. Pacati, Phys. Rep. {\bf 226}, 1 (1993).
%
\bibitem{Ud1}
J.M. Ud\'{\i}as, P. Sarriguren, E. Moya de Guerra, E. Garrido, and 
J.A. Caballero, 
Phys. Rev. C {\bf 48}, 2731 (1993);
C {\bf 51}, 3246 (1995)
%
\bibitem{meucci1}
A. Meucci, C. Giusti, and F.D. Pacati, 
 Phys. Rev. C {\bf  64}, 014604 (2001).
%
\bibitem{amaro05}
J.~E. Amaro, {\it et al.}, 
 Phys. Rev. C {\bf 71}, 015501 (2005). 
%
\bibitem{ee}
A. Meucci, F. Capuzzi, C. Giusti, and F.D. Pacati,
 Phys. Rev. C {\bf 67}, 054601 (2003).
%
\bibitem{confee}
  A.~Meucci, J.A.~Caballero, C.~Giusti, F.D.~Pacati, and J.M.~Ud\'{\i}as,
Phys. Rev. C {\bf 80}, 024605 (2009).
%
\bibitem{cab10}
  J.~A.~Caballero, M.~C.~Martinez, J.~L.~Herraiz, and J.~M.~Ud\'{\i}as,
  Phys. Lett.  {\bf B688}, 250 (2010).
%
\bibitem{Chiara03}
C. Maieron, M.~C. Mart\'{\i}nez, J.~A. Caballero, and J.~M. Ud\'{\i}as,
Phys. Rev. C {\bf 68}, 048501 (2003).
%
\bibitem{cab1}
J.~A. Caballero, Phys. Rev. C {\bf 74}, 015502 (2006).
%
\bibitem{cc}
A. Meucci, C. Giusti, and F.~D. Pacati, 
Nucl. Phys. {\bf A739}, 277 (2004).
%
\bibitem{hori}
Y. Horikawa, F. Lenz, and N.C. Mukhopadhyay,
%Y. Horikawa,   {\it et al.},
Phys. Rev. C {\bf 22}, 1680 (1980).
%
\bibitem{cab07}
 J.~A.~Caballero, J.~E.~Amaro, M.~B.~Barbaro, T.~W.~Donnelly, and J.~M.~Ud\'{\i}as,
  Phys. Lett. {\bf B653}, 366 (2007).
%
\bibitem{eenr} 
F. Capuzzi, C. Giusti, and F.D. Pacati, 
Nucl. Phys. {\bf A524}, 681 (1991). 
%
\bibitem{eeann} 
F. Capuzzi, C. Giusti, F.D. Pacati, and D.N. Kadrev, 
Annals Phys. {\bf 317}, 492 (2005).
%
\bibitem{Meucci:2005pk}
  A.~Meucci, C.~Giusti, F.~D.~Pacati,
  %``Relativistic Green's function approach to parity-violating quasielastic
  % electron scattering,''
  Nucl.\ Phys.\  {\bf A756}, 359 (2005).
%
\bibitem{Meucci11}
A. Meucci, J.~A. Caballero, C. Giusti, and J.M. Ud\'{\i}as,
Phys. Rev. C {\bf 83}, 064614 (2011).
%
%
\bibitem{Tina10}
T. Leitner and U. Mosel,
Phys. Rev. C {\bf 81}, 064614 (2010).
%
\bibitem{cab05}
J.~A. Caballero, J.~E. Amaro, M.~B. Barbaro,  T.~W. Donnelly, C. Maieron, and  
J.~J.~M.Ud\'{\i}as,
Phys. Rev. Lett.  {\bf 95}, 252502 (2005).
%
\bibitem{amaro07b}
J.~E. Amaro, M.~B. Barbaro, J.~A. Caballero, and  T.~W. Donnelly, 
Phys. Rev. Lett.  {\bf 98}, 242501 (2007).
%
\bibitem{Meucci06}
  A.~Meucci, C.~Giusti, and F.D.~Pacati,
  Nucl.\ Phys.\  {\bf A773}, (2006) 250.
%
\bibitem{Meucci:2006cx}
  A.~Meucci, C.~Giusti and F.~D.~Pacati,
  Acta Phys. Polon.  B {\bf 37}, 2279 (2006);
  Acta Phys. Polon.  B {\bf 40}, 2579 (2009). 
%
\bibitem{bound}
B.D.~Serot and J.D.~Walecka, Adv.\ Nucl.\ Phys.\ {\bf 16}, 1 (1986);
M.M. Sharma, M.A. Nagarajan, and P. Ring,
Phys. Lett. {\bf B312}, 377 (1993).
%
%
\bibitem{chc} 
   E.D. Cooper, S. Hama, B.C. Clark, and R.L. Mercer,
 Phys. Rev. C {\bf 47}, 297 (1993).
%
\bibitem{Cris06} 
M.C.~Martinez, ,   {\it et al.},
Phys. Rev. C {\bf 73}, 024607 (2006).

\end{thebibliography}
\end{document}